# Players' Perception of Bugs and Glitches in Video Games: An Exploratory Study

Jessica Backus

Purdue University, Department of Computer Graphics Technology

backus@purdue.edu

**Abstract**
The goal of this exploratory research is to investigate how glitches and "bugs" within video games affect a player's overall experience. The severity or frequency of bugs, as well as the nature of the bugs present, could influence how the players perceive these interactions. Another factor is the player's personality because this will affect their motivations for playing certain games as well as how they react to bugs within these games. Glitches and bugs are framed as a negative aspect within games, but create the potential for enjoyable experiences, despite being unexpected. To explore this hypothesis, I observed some glitches within recorded gameplay via YouTube and Twitch livestream VODs and analyzed the streamer's reaction, as well as the audience's. I also conducted semi-structured interviews with gamers with the goal of learning more about that player's personality and attitudes towards bugs in the games they play. I concluded that the types of bugs matter less to the players than how frequently they occur, the context they occur, and the outcome of them.

## 1 Introduction

Glitches or bugs are unavoidable when creating games, no matter how skilled the developers are. Despite the prevalence of glitches within games, little research has been done on exploring the effects that these glitches have on the player's overall experience from a subjective standpoint. The negative connotation of errors and bugs leads people to think that they must create only a negative impact on the player's experience. It was an experience that was never intended by the developers; therefore, it must be bad. However, as shown by the vast amount of gaming videos featuring glitches and bugs in a variety of games, players have engaged with these bugs with purpose and enjoyment. They perhaps have grown accustomed to having these bugs within their games and therefore have created communities or certain practices within gaming communities that center around these bugs. To learn more about this phenomenon, I first aim to learn more about bugs in general, learning about the different types of bugs. Then to learn how these different bugs are perceived by players, I ask:

1. What are players' attitudes towards glitches and bugs?

This first research question will allow me to gauge how players' opinions on common kinds of bugs vary between different players. I also want to know how the developers affect this problem as well, specifically:

2. How does the Developer's response to bugs affect the player's opinions on the game?

While bugs can happen unintentionally, they can also be triggered intentionally by seasoned players. Hence, I also ask:

3. How do intentional vs. unintentional bugs affect the player's experience, as well as other players in the case of multiplayer games?

## 2 Background
The existing research involving bugs in games has focused mainly on what kinds of bugs exist [8], which ones to fix and when [6], and the outcomes of these bugs [7, 11]. To help analyze the bugs that I find in my observations as well as what is discussed in my interviews, I wanted to learn what kinds of bugs are common. Lewis et al. [8] breaks down several types of bugs they found in their survey of video recordings of glitches in games on YouTube. The taxonomy they use for categorizing these glitches aligns with the language used in game development, so to summarize these categories, I group them based on similarity and give them a more simplified name, for easier comprehension and comparison.

*2.1 Types of Bugs*
Lewis et al. [8] discusses in depth the different bug types that they witnessed, including some examples and specific games those bugs were in. As discussed in [2], I categorized these bugs, focusing on four elements: the player, the game objects, actions/interactions, and the game systems. Bugs can affect different aspects of a game in very complex ways, but by separating the game into these components, we can quickly understand which features of the game contain the errors.

*2.2 Glitches/Bugs and Unfairness*
During my background research, I came across several papers that discuss bugs and glitches in various contexts, including online multiplayer games, single-player games, and esports. Among all these papers was the concept of unfairness, or cheating. From this, the main thing that I took away from these papers was that the concept of unfairness depends on two things: context and opinion. In the context of Esports, Irwin and Naweed [7] explore the different behaviors of players in Esports tournaments, including the exploitation of bugs. There were instances of players exploiting known bugs in order to get an advantage in the game, *Counter-Strike: Global Offensive*, that they would use to win matches, or in some cases, the entire tournament. Arguments for the use of bug exploitation say that the players were simply utilizing all parts of the game and explored every opportunity presented and did not alter the game in any way. Others claim that any scenario in which one player has an advantage over another is a form of cheating. Consalvo and Vazquez [4] also state that "Most players agree that cheating is anything that gives a player an unfair advantage." Like with any other sport, there is a lot of debate about the rules and where they must draw the line at what is considered cheating. Esports also has the added consequence of prize earnings, often in the form of monetary winnings. This adds even more pressure to the debate, considering there is much more to be won than just a game.

But even outside the context of Esports, players often share the same ethics about bug exploitation in multiplayer games. Consalvo [3] found that players have different ideas of cheating with different degrees of "allowances" or what they would allow without considering it cheating. In multiplayer games, it was said that "players defined cheating as only existing in relation to another player", meaning that there had to be some negative effect on another individual for it to qualify as cheating. For these players, any sort of advantage, whether from bug exploitation, hacks, or other means of altering gameplay, was not considered cheating if in the context of single-player games. There are other players, however, that believe that cheating involves relying on anything outside of the game to progress or altering the gameplay on any level. This includes walkthroughs, assistance from a friend, or the use of "cheat codes", no matter if the game was single- or multiplayer.

A study by Švelch [13] also discusses bug exploitation, but in the interesting context of a game with microtransactions. In one scenario, bugs could be utilized to bypass the use of microtransactions AND give a player an unfair advantage over others. This was quickly deemed a bug by the developers, and they patched it. In another game by the same developers, there was another bug that allowed players to gain in-game materials to help them level up but required slightly less time and effort than normal. A microtransaction available would *instantly* give a player these materials, so because this microtransaction was not directly superseded, it was deemed a *feature* by the developers. However, another important thing to note is that the first bug went against the developer's intention and the players' expectations for normal gameplay, while the latter only helped players "farm" materials in the game. This distinction is important because bugs like the first are more easily measured and more agreed upon by all stakeholders that it is indeed a bug, not a feature, unlike the second which was more ambiguous/subjective. It then begs the question if the microtransactions that afford the player instant benefits over others who do not purchase the microtransactions, is this, in a way, cheating? As mentioned previously, there are some players that would consider this cheating, since it falls outside of "normal" gameplay [3].

*2.3 Motivations for Bug and Glitch Exploitation*
While cheating is the primary reason why players may view bug exploitation in a negative light, there are positive perspectives that motivate players to learn about and utilize bugs. One of the most popular examples of this is "Speed-running." Speed-running is when a player goes through a game with the goal of completing it with the quickest time possible. There are many subcategories of speed-running including "Any%", "100%", "no-glitches", or runs without using a specific glitch, to name a few. Speed-running as a method of playing a game is unique because players are "sidestepping the intended and expected rules of playing the game as defined by the designer in order to redefine the goals of the game for their own purposes as a community" [5]. This means that players are treating the game as a dynamic entity, rather than a rigid body with set "laws" determined by the code/developers. Not only do they encourage this bending of rules, they seek out glitches to use in their speedruns and share them with their community.

This concept of finding glitches and creating a community around them covers two needs discussed by Przybylski et al. [9] using self-determination theory (SDT); the competence need (specifically "Mastery of Controls") and the relatedness need. Mastery of controls is a key factor for some players and their motivations for playing certain games. Player types like "achievers" discussed by Aarseth [1] would especially use this as their motivation to play, since they are most motivated to accomplish the most in a game. The "socializer" player type [1] align with the relatedness need in SDT since they are most motivated by the interactions they have with others while playing, seeking community and welcoming cooperative challenges. A third player type would also fit within this paradigm: the "explorer" type. Explorers are most motivated to uncover all components of a game, including its secrets or "Easter eggs" and any glitches they may uncover. These player types are essential for speedrunning communities since they would be the type to discover and share the glitches/bugs they uncover.

**3 Methods**
To learn firsthand about the opinions that players have about the glitches and bugs in their games, I conducted interviews with 6 players, and observed 2 livestream recordings of two different games from the same streamer. Interviewing players allowed me to learn about their motivations for player certain kinds of games and the kinds of glitches and bugs that they had personally come across. The observations allowed me to observe glitches and bugs "in action" as well as the reactions from the streamer and their audience (when chat was included in the recording). Viewing these errors and reactions in context helped to reveal some "big picture" insights that otherwise could be overlooked.

*3.1 Observations*
With the nature of this research topic, it would be ideal to observe a bug or glitch occur in real time, but it is hard due to its "randomness" (unless the player is very well-versed with the game and is intentionally trying to create a known glitch, such as in speed-running). As a member of several gaming communities, I watch many livestreams and videos from my favorite creators. During this research project, I was fortunate enough to tune into one livestream where the streamer witnessed a glitch happen to his friend and I witnessed both the streamer's and the audience's reactions live. However, due to being a participant myself at the time, I did not take any field notes. To fully document this observation, I returned after the livestream and watched the "video on demand" and took 30 minutes' worth of field notes, including the moments of a glitch occurring. For the second observation, I found a livestream from the same streamer featuring a popular glitch clip of his and recorded 30 minutes of this play session around the time the glitch occurred. Unfortunately, the audience chat was not available for this second observation, so only the streamer and his teammate's reactions were recorded.

*3.2 Interviews*

As stated above, I interviewed 6 people, all with several years of playing experience (4-15+ years), spanning several genres of games and on different platforms. My method for recruiting these interviewees was a combination of convenience and snowball sampling. I knew several of these interviewees were avid video game players and would provide useful insights for my research, and they suggested I interview some of their friends as well. Each interview lasted between 30 minutes to 1 hour and was done over a video call via Zoom, which was also used to export a full transcription of each interview. During the semi-structured interviews, I also showed the interviewees two gaming clips, one featuring a bug and the other showcasing "unexpected features" that initially appear as a bug to gauge their reactions to both. A full interview protocol can be viewed [in the appendix.](#)

*3.3 Thematic Analysis*

After collecting field notes from my observations and transcripts from my interviews, I coded my data following methods listed in [10] using the online qualitative analysis platform *Dovetail*. The first pass of coding focused on what topics were discussed in the quotes, such as "multiplayer games", "single-player games", and "bug types", as well as more abstract topics like "exploiting motivations", "attitudes toward glitches", and "annoying." After this first pass, I then did a second pass of "versus" coding, focusing on "positive" and "negative" views of bugs/glitches and their outcomes. However, in this process, there was a third category that was more in the middle of positive and negative that wasn't exactly neutral. It was more of a conditional scenario that I dubbed "Frequency." I explain this in detail in the following section. After this second pass of coding, I generated themes and insights based on the patterns I found throughout my interviews and observations.

**4 Results**

The first theme that I generated stemmed from the "negative" code, focusing on bug exploiting in the multiplayer context. The players that I interviewed had opinions that agree with the results from Consalvo [3], that bug exploiting is negative when it affects another player, and that their concept of cheating is inherently social. Some quotes from the interviewees that display this line of thinking are:

*"If you're in a competitive setup, right, and then you're using a bug exploit and you're being very obvious about it or, you know, you're just being a bad player throughout."*

*"The developers want you to experience one thing. You're moving away from that experience and you're spoiling the other person's experience."*

*"When people exploit bugs and glitches in our multiplayer games on especially in esports tournaments then that becomes a bit problematic."*

*"It's interesting and it's fun, but the moment it starts to impact the other person's game, then that kind of takes away from the whole game experience."*

*"They were not trying to have fun with the game, they were just trying to be like 'oh I want to get to radiant as fast as I can so I'll use this bug to kill as many people as possible to be the best ranked player in my region.'"* \*

\*This quote is especially interesting because this player is aware of the motivations behind the exploitation of bugs in his game.

The theme created from this insight is **"Exploitation of bugs negatively affect other players."**

This means that when a bug is used in a multiplayer context intentionally, players see it as something that negatively effects all player's experience when used to gain an advantage, and often equate it to cheating.

But what about glitches in multiplayer that aren't used to gain an advantage, or are done unintentionally? This is where my "positive" code focused. There are many videos available on sites like YouTube of players finding bugs within their games, some of them showing how to perform the glitch, others simply displaying it due to its unexpectedness and the humor surrounding it. One of my observations is an example of the latter. In this observation, the streamer [12] learns about a glitch from one of his teammates and seems to successfully perform the glitch, with some unexpected side effects. In *Call of Duty: Warzone*, these players found a work around to put a sentry turret on top of a vehicle (which should not be possible). This causes the truck "bug out" and start clipping into the ground, becoming nearly impossible to drive and it results in one of the streamer's teammates being eliminated. The unexpected and hectic nature of the bug caused by the glitch was viewed as very comedic:

*"Watching it in third person was so funny. We gotta do that again."*

*"It's been so long since I've laughed that hard…"*

The comments of the YouTube upload also found this moment to be particularly enjoyable:

*"44:00 the legendary clip."*

*"45:30 best part of the vid 😂"*

Even though the glitch was intentional, the bug caused afterward was not, which created a very memorable moment for the streamer, teammates, and audience. Interviewees also mentioned that sometimes moments like these can be enjoyable.

*"All the games I've played, the bugs and glitches were always hilarious and fun to me. It's always hilarious and fun to me."*

*"I mean, I think the bugs kind of add to the fun of the game."*

One interviewee described their own moment when they experienced a bug in the game *Phasmophobia*, a ghost hunting game:

*"So, they're just laughing, and we're supposed to be scared of the ghost. But instead, it's looking around and we're all standing there laughing at it."*

The ghost in question was stuck on a piece of furniture and could not move or chase the players (which is a case of "dumb AI" from [8]), which in that moment, broke the immersion and was humorous, rather than scary (which this moment was supposed to be). These unexpected breaks in the games immersion or what goes against the player's expectations of the game leads to moments of hilarity and rather ruining their fun, it adds to it. The same interviewee stated:

*"It was an interesting interaction. We all had a good laugh about it and by the time it was over, the game actually got back to its proper state."*

*"In fact, I'd say that it's something that elevated it. Right, but obviously I wouldn't want it happening again and again."*

This leads to my next theme of "**Unexpected bugs in multiplayer create fun moments.**"

However, the last point from the previous quote was a pattern I found throughout several interviewees.

*"[The] first 15 times it's fun and then it gets it gets really annoying."*

*"Okay, initially it's funny then it becomes frustrating because you are completely immobile after a while. And then… you are like, okay, I'm gonna lose my progress."*

*"Initially, like when I experienced it for the first time or the first couple of times, I think I just sort of laughed it off…after a while, I think it starts to get really frustrating."*

So, this brings about a very interesting insight that the perception of these moments depends on frequency. When they occur in small amounts, at first, they are seen as humorous and enjoyable. Conversely, if these bugs were to continue to happen over and over again, they quickly become frustrating. Without numerical data to show this, I imagine that this is like an exponential relationship where the number of times the bug happens increases, the more frustrated the player becomes. Thus, my next theme: **"Frequent, reoccurring bugs create frustration."**

But this brings about new questions, such as "what player types would find these reoccurring bugs less frustrating?" or perhaps, "In what *contexts* would these reoccurring bugs be less frustrating?" One hypothesis I have is that the presence of friends during gameplay can create less tension for players and increase their tolerance of bugs. The second observation I did was about a game called *The Long Drive"* which is a $16 post-apocalyptic driving game on Steam. Throughout my thirty-minute observation, I witnessed at least 5 different instances of bugs or glitches [11]. Despite all these bugs transpiring in such a short amount of time, the streamer, his friend, and the audience all seemed to find enjoyment in the chaos. Some moments from this observation:

Streamer is laughing: "*What f\*\*\* happened?*"

Chat: *"The wheeze lmao"*

Chat: "*I didn't even know that could happen KEKW"\**
\*KEKW is a Twitch emote of a man laughing

Chat*: "This stream has been the most hilarious s\*\*\* so far."*

Even though the bugs present took some extra effort on behalf of the players to fix (they had to leave the game and rejoin to fix them), they "laughed it off" and continued their game. One interviewee also shared this idea of these situations being better with friends:

*"I mean, in a group setting, it's kind of fun, like when you're playing with your friends and stuff."*

This idea of having friends present during unexpected bugs creates the theme: **"Socialization during unexpected bugs positively alters their perception."**

Following this idea of socialization during unintentional bugs, what happens when no one else is around? As mentioned before, there is the situation of bugs appearing in single-player games intentionally with speed-running. However, when a player is playing a single-player game with no intentions of using a known bug or glitch, their immersion is broken when a bug occurs. Several interviewees that play single-player games say that their main motivations for playing these kinds of games is the story, the game world, and the immersion they experience due to these components. Having bugs occur in these contexts can "ruin" these moments, or the game all together. Some interviewees gave some examples:

*"I was playing Dark Souls 3. I got towards the end. I just had this final segment that I needed to do to complete the game. Just before I could do that, my safe file was corrupted, and I lost all my progress."*

*"I've experienced a lot when I was playing. It was pretty f\*\*\*ing annoying because this was like, a game-breaking bug. So, once you go down you just cannot go up, like, at all… so you need to restart the game. So, this happens a lot in like middle of levels. So, I need to play that whole level again. It was pretty annoying."*

*"Like if there's a glitch in the boss fight… that would kind of ruin the whole experience."*

Because the immersion of their single-player game is ruined and they don't have the socialization aspect as with multiplayer games, these unintentional bugs are immediately an unpleasant experience for these players. For this specific context, I made the theme: **"Unintentional bugs in immersive single-player games are innately negative."**

The last major topic that was discussed (and relates back to one of my research questions) is the actions of developers in response to bugs. Primarily, I asked interviewees how they felt when developers patch certain bugs and who is responsible in the circumstance of bug exploitation, the players, or the developers? For the first situation, interviewees said:

*"In my opinion, like for example, [the bugs] the Spiderman Devs took out, now you cannot do it anymore and that is kind of, I don't know, that kind of makes the game kind of boring."*

*"[The bug] made the game much more fun, in my opinion. I mean game breaking bugs you do need to fix that because obviously that's f***ing breaks the game."*

*"The most successful games don't have a lot of bugs at all."*

*"It depends on how long the bug has been out. Like it's just been 1, 2 days and I'm like, okay, the devs are gonna fix this. If it's been like a week and the bug is still there, I would just be like, oh, I don't really care about playing this anymore."*

Overall, they had mixed opinions about developers patching bugs, but this could go back into the fact that the frequency and the context alters how these players feel about the bugs to begin with. For competitive multiplayer games or severe bugs, it makes sense that players prefer if these bugs are patched vs. "minor" ones in non-competitive contexts.

When it comes to exploiting bugs and reporting them, players also have some conflicted feelings about where responsibility lies. When asked about this, some players said that the responsibility lies with the developers:

*"It should not be left to the public in general because…they are not there to, you know, find the bugs for you."*

*"But that it is something that the developer should be looking into. Just to make sure that doesn't happen a lot."*

*"I think it's the developer's responsibility to seek out whether or not those bugs are occurring. And then patch them up accordingly."*

*"I think it's their responsibility. I mean, the developers because we paid them."*

However, some said that being a good member of the community meant helping report bugs:
*"I mean, you're just helping the whole community like making the game better."*

*"So, it should be your duty as a good player and to make a good gaming community to report it."*

One participant was more in the middle:

*"Depends on what game it is. If it's a multiplayer game, I would say both [the developers and the players]."*

In general, players agree that developers are the primary ones responsible in finding and fixing bugs (especially when it comes to the issue of bug exploitation), but they also agree that by reporting bugs, they are helping their community. However, which bugs to patch and when is less clear. Bugs that have less impact on the overall gameplay and create fun moments as the interviewees said could be considered "low priority" so that developers can focus on higher priority issues. This creates my last theme: **"When and how developers respond to bugs positively or negatively alter player's opinions."**

## 5 Limitations
With only two observations and six interviews, the primary limitation that I have in my research is my small sample size. Also, my method of recruiting interviewees was limited to people that I knew (convenience sampling) and other people that they knew and recommended to me (snowball sampling). If I were to do more work with this in the future, I would do criterion sampling to ensure that candidates are more "random", but still meet my needs. Also, all of my interviewees were graduate students at the same college, in the same department, all around the same age. If I were focusing on opinions from gamers within the Computer Graphics Technology department, my method of recruitment would be fine, but this means that for now, it's hard to generalize my findings.

My method of obtaining observations also imparts some bias because I am a member of the online community that I observed. While this does provide some helpful context that may not be available to someone outside of this community, it does impart some added bias in my data collection and interpretation. These observations only cover two games as well, so observing more glitches in more games could provide more insights.

**6 Discussion**
Players have a good understanding of the game development process and what goes into creating/maintaining a game. They recognize what the developers' intentions are for a game and that shapes their expectations for the game and its contents. Whenever something in their game contradicts these expectations, the believe that this is a bug or a glitch. What determines the player's reaction to these bugs/glitches is **the context**. Whether it is a single-player game or competitive multiplayer, or if the glitch was found intentionally or unintentionally, the players will have different reactions. Another large factor is **how frequently** they encounter bugs. The presence of frequent bugs quickly (and negatively) alters the players opinion of the game. Lastly, what affects the player's opinion of the bug/glitch is **the outcome**. Does the bug fix itself? Or does it entirely ruin the player's game? To demonstrate my understanding of how bugs and glitches affect a player's game overall, I created a concept map, shown below in Figure 1.

Players agree that there are times when bugs are not only okay to have in their game, but when they are beneficial. Going back to the concept of speedrunning [5], interviewees were familiar with this type of gameplay, even when they themselves did not participate in it:

*"So, if people post videos of, you know, doing speed runs using bug exploits or, you know, playing the game using some glitch as a hack. I think that's completely fine."*

*"Yeah, it's… you are not, you know, you are just breaking the game for your own fun and it's hilarious sometimes as well and also it helps speed runners to perfect their speed runs so that's fine for me."*

*"Speedrunning is always fun."*

In this context, the bugs are exploited intentionally such that the player receives benefits towards their own personal goals, rather than goals set in the game or advantages over other players. This is why players like my interviewees, the streamer I observed, and players like the ones described in [3] would not see a problem with utilizing glitches in this context, since they would not be hindering the gameplay of others (and therefore would not consider it cheating).

The other ways that bugs can be perceived as a positive align with my themes: "**Unexpected bugs in multiplayer create fun moments**" and **"Socialization during unexpected bugs positively alters their perception."**

Players find glitches fun in certain contexts, so long as they believe they are not hindering someone else's experience, aren't morally wrong by breaking rules where the consequences are more dire (Esports), or ruining their own gameplay. They are likely to enjoy these moments more when they can share them with friends or with others online, especially the "socializer" player type [1].

On the other hand, players perceive bugs negatively when they impact other players, aligning with my theme of **"Exploitation of bugs negatively affect other players."** Other negative perspectives that align with my themes include: **"Frequent, reoccurring bugs create frustration"** and **"Unintentional bugs in immersive single-player games are innately negative."**

Players find glitches and bugs annoying when they are the ones at a disadvantage to bug exploitation by other players, when bugs ruin their immersion in single-player story-based games, and when the bugs completely break their game or the flow of their game. It's possible that one minor bug that rights itself in a single-player game can be overlooked by the players but going back to the "frequency" point discussed before, if it were to happen several times, or if the bug did not right itself, this would quickly impact the player's experience. This is why for this specific context it's important that less or no major bugs make it into the final game.

Lastly, "**when and how developers respond to bugs positively or negatively alter player's opinions.**" If the developers patch a "fun" bug, the players view the game and the developers as less fun to engage with, and when "bad" bugs are not patched, the players believe the developers are not being responsible by fixing their product and are negatively impacting their gameplay. If they even go a step beyond and deem that bug a feature, players criticize the developer's and say that it was a cop-out for
them [13].

**7 Conclusion**
By interviewing and observing players, I was able to find certain patterns in their behaviors and opinions centering around bugs and glitches in their games. Even though I am a member of a few gaming communities, there were some findings that I found surprising, but in hindsight, make logical sense.

I've learned that it's not the types of bugs that matter, it's the context, frequency, and outcome that determines their perception. Obviously, large, game-breaking bugs that ruin the game entirely are still considered negative, no matter the context. By keeping this in mind, developers can make better decisions about what bugs to patch first, which ones to patch *at all*, and which ones they can alter such that they become more of a feature that improves gameplay for their players. Instead of trying to make a "perfect" game with no bugs at all (which is impossible to achieve), developers can embrace the smaller bugs they do have while eliminating or lessening larger ones, depending on their goals and expectations for their games. The best way to determine where to focus their attentions is to learn about the player types, behaviors, and goals of their player base so they can create a game that is a positive experience for them specifically, not just what they believe that their players want or their ideas of what makes a good game.

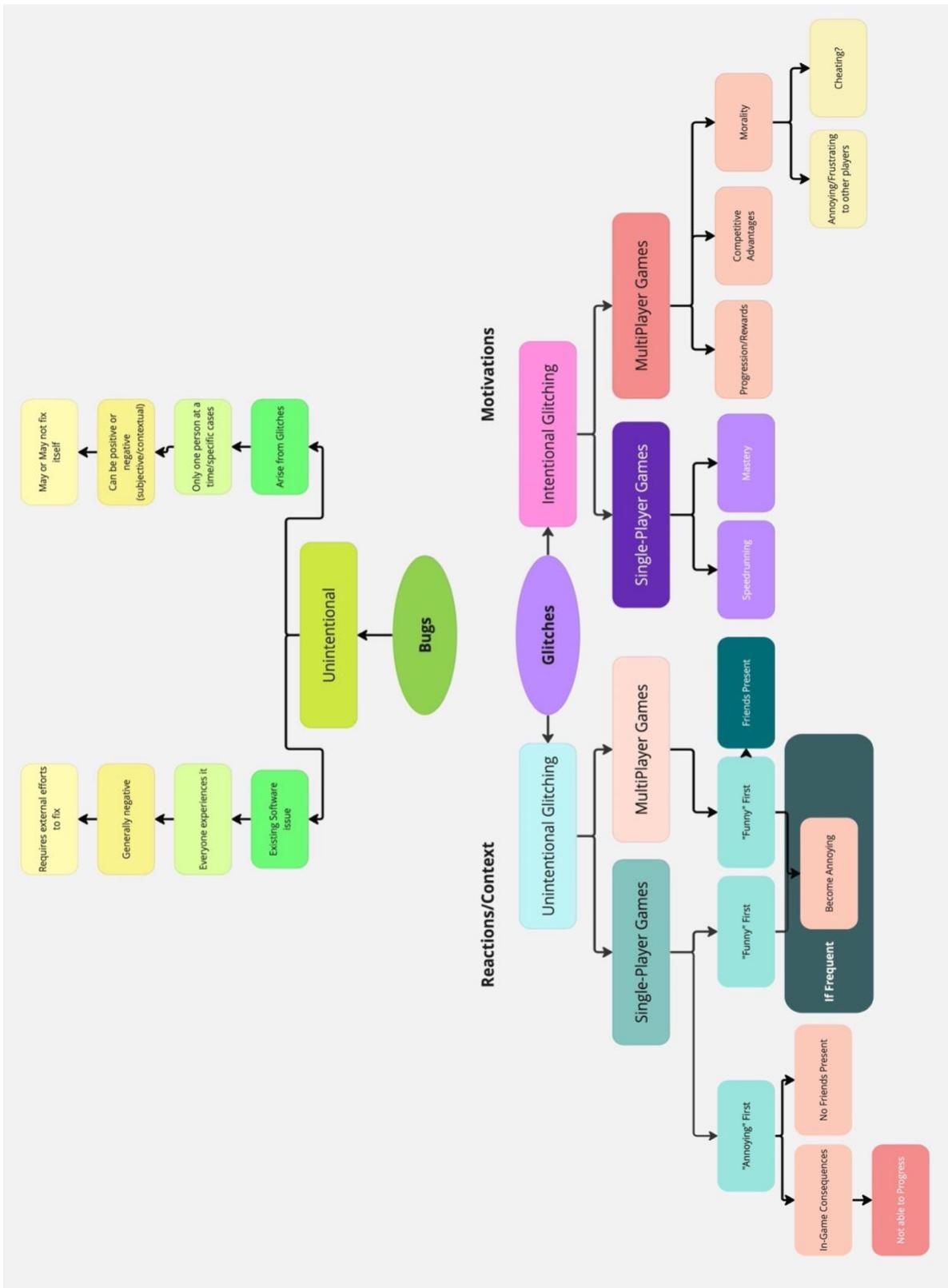

Figure 1: Concept Map of the perception of bugs and glitches

# 9 Appendix

## 9.1 Interview Protocol

"Thanks again for agreeing to be a participant in my research! I asked you to be a participant because you have experience with gaming and are an active player for various kinds of games. To begin, I will ask some simple questions about yourself, to get a better idea of what kind of player you are."

*Personal Background*

- How long have you been playing video games?

- What are your favorite games, or favorite genre?
- How often do you play games? Estimated hours per day/week/month?
- How has your average time playing changed over time?
- What prevents you from playing games?
- What do you typically play on? Console, PC, mobile?
- What are your primary motivations for playing games?

"Okay! So, because I'm researching bugs and glitches within games, I have some questions about your own experience with those things. Firstly…"

*Glitches/Bugs*
- How would you define a "bug" in a game vs. a "glitch"?
- Throughout your gaming career, have you experienced glitches or bugs in your games?
- If so, what did these look like?
- How "severe" were they? Did they "break" the game, or just cause some interesting interactions? Or even improve the experience?
- Would you say that the bug(s) greatly impacted your experience?
- What game was it? When?
- Has the bug been patched (fixed)? Why do you think so, if yes?
- Have you ever intentionally sought out bugs or tried to "glitch" the game? Why or why not?
- If yes, do you experiment (try to find a bug on your own), or look within communities for instructions?
- Where do you look aka what communities do you participate in? These can be online or in-person.
- What are your personal thoughts on bug exploits? Have you ever exploited a bug? Why or why not?
- Do you think it is the responsibility of players to not exploit bugs and rather report them when they happen? Or is it the responsibility of the developers to act swiftly and/or make sure they do not happen in the first place?

"Now I'm going to show you a few videos of examples of glitches/bugs that I found online. Feel free to comment on these at any point."

*Post-Video observation*
- Have you seen this bug/bugs like this before?
- Have you played a game where a similar bug happened to you? What was your reaction?
- Did this/these bug(s) alter your experience? Positively, negatively?
- Would seeing other players exploit this bug make you less or more motivated to play? Or are you indifferent?

Do you have any additional comments? Or any questions? Thanks for your help!

12